\documentclass[10pt]{jfm}
\usepackage{physics}
\usepackage[usenames, dvipsnames]{color}
\usepackage{ar}
\usepackage{amsmath,color,graphicx,amssymb}
\usepackage[font=small]{caption}
\usepackage{amsbsy}
\usepackage{latexsym}
\usepackage[normalem]{ulem}
\usepackage[english]{babel}
\usepackage{ifsym}
\usepackage{bbding}
\usepackage{wasysym}
\usepackage{latexsym}
\usepackage{epsfig} 
\usepackage{epsf,psfig}
\usepackage{epstopdf}
\usepackage{multirow}
\usepackage{array}
\usepackage{float}
\usepackage[english]{babel}
\usepackage{verbatim}
\usepackage{soul}
\usepackage[toc,page]{appendix}
\usepackage{setspace}
\usepackage{threeparttable}
\usepackage{booktabs,array}
\usepackage{color}
\usepackage{mathtools}
\usepackage{siunitx}
\usepackage{setspace}
\usepackage[numbers]{natbib}
\usepackage{authblk}
\usepackage{gensymb}
\usepackage[section]{placeins}
\usepackage{caption}
\definecolor{lightblue}{RGB}{48, 89, 155}
\definecolor{black}{RGB}{0, 0, 0}

\usepackage{float}
\def\bibfont{\footnotesize}
\usepackage{trimclip}

\begin{document}

\newtheorem{lemma}{Lemma}
\newtheorem{corollary}{Corollary}

\shorttitle{High-efficiency self-organizing hydrofoil schools} 
\shortauthor{Tianjun Han et al.} 

\title{Tailoring formations of self-organizing hydrofoil schools towards high-efficiency}


\author
 {
Tianjun Han\aff{1}
  \corresp{\email{tih216@lehigh.edu}},
  Amin Mivehchi\aff{1},
  Seyedali Seyedmirzaei Sarraf\aff{1},
  and Keith W. Moored\aff{1}
  }

\affiliation
{
\aff{1}
Department of Mechanical Engineering and Mechanics, Lehigh University, Bethlehem, PA 18015, USA
}

\graphicspath{{Figures/}}
\maketitle

\begin{abstract}
We present new unconstrained simulations and constrained experiments of a pair of pitching hydrofoils in a leader-follower in-line arrangement. Free-swimming simulations of foils with \textit{matched} pitching amplitudes show self-organization into stable formations at a constant gap distance without any control. Leading-edge separation on the follower foil plays a crucial role in creating these formations by acting as an additional dynamic drag source on the follower, which depends on the gap spacing and phase synchronization. Over a wide range of phase synchronization, amplitude, and Lighthill number typical of biology, we discover that the stable gap distance scales with the actual wake wavelength of an isolated foil \textit{rather than} the nominal wake wavelength. A scaling law for the actual wake wavelength is derived and shown to collapse data across a wide Reynolds number range of $200 \leq Re < \infty$. Additionally, in both simulations and experiments \textit{mismatched} foil amplitudes are discovered to increase the efficiency of hydrofoil schools by 70\% while maintaining a stable formation without closed-loop control. This occurs by (1) increasing the stable gap distance between foils such that they are pushed into a high-efficiency zone and (2) raising the level of efficiency in these zones. This study bridges the gap between constrained and unconstrained studies of in-line schooling by showing that constrained-foil measurements can be used as a map of the potential efficiency benefits of schooling. These findings can aid in the design of high-efficiency bio-robot schools and may provide insights into the energetics and behaviour of fish schools.

\end{abstract}

\section{Introduction}
Aquatic animals school for many reasons from protection against predators and enhanced foraging to increased socialization and saving energy while swimming \citep{Weihs1973,  Marras2015a, Ashraf2017, Li2020}. In fact, this energy savings, derived from the hydrodynamic interactions that occur in schools, has been confirmed through direct energy measurements of live fish \citep{Zhang2023, Zhang2024a}. These energy benefits can be observed not just among fish in a school, but also between oscillating foils and fish \citep{Harvey2022a,Thandiackal2023}, and among oscillating foils \citep{Ristroph2008, Lin2019, Lagopoulos2020, Kurt2020, Kurt2021, Lin2022}, which supports the idea that schooling hydrofoils can serve as a simple model of schooling fish and can also be used to propel high-performance underwater biorobots.


In-line formations are canonical schooling arrangements where swimmers are aligned in a leader-follower pattern with followers directly downstream of a leader. Most previous research on in-line schooling has focused on so-called ``constrained" or ``tethered" foils that have fixed relative spacings. In-line formations of constrained foils show some of the largest hydrodynamic benefits of schooling, where collective thrust and efficiency enhancements have been measured of more than 50\% as compared to an isolated swimmer \citep{Akhtar2007, Ristroph2008, Boschitsch2014, Muscutt2017, Kurt2018, alaminos2020aerodynamics,  Wang2021, alaminos2021propulsion, arranz2022flow, Han2022a,  baddoo2023generalization}. These hydrodynamic benefits arise when the wake vortices shed from a leader constructively interact with the leading-edge suction of a follower, and thereby increase its thrust, primarily driven by a modification of its instantaneous angle of attack from the impinging vortices \citep{Akhtar2007, Boschitsch2014, Muscutt2017}. When a leader's wake vortex impinges on a follower it also induces the formation of a leading-edge vortex that pairs with the impinging vortex to form either a coherent or a branched wake mode where thrust is maximized or power consumption is minimized, respectively \citep{Boschitsch2014,Kurt2018}. Additionally, band structures of high follower efficiency \cite[]{Boschitsch2014,Kurt2018} reveal that the optimal phase difference or synchrony between a leader and follower follows a linear relationship with their streamwise spacing, which arises from the optimal synchronization of the follower's motion to the impinging vortex. While the collective performance is mostly driven by the followers' performance,  in a compact formation a follower can also have a significant upstream influence and enhance a leader's added mass thrust by providing a blockage effect \citep{Saadat2021, Pan2022, Kelly2023}. Therefore, the optimal collective efficiency is achieved for relative spacings of less than one chord, and with a properly tuned phase synchrony for a given spacing.

While constrained foil studies provide essential understanding of schooling mechanisms and their \textit{potential} benefits, their foils are mostly in \textit{out-of-equilibrium} conditions, meaning that there are thrust imbalances among them that if they were freely swimming, as in real schools, would cause their relative spacing to change and in turn alter their hydrodynamic interactions and benefits. Studies of \textit{unconstrained} in-line foils able to freely swim in the streamwise direction have shown that streamwise stable spacings arise from their hydrodynamic interactions alone \citep{Becker2015, Newbolt2019, Heydari2021, newbolt2024flow}. These fluid-mediated formations have a gap distance between foils that is nearly constant, indicating that there is thrust balancing between them, and they are \textit{stable} since small perturbations from these formations will cause the foils to return to them.  These stable spacings were discovered to vary linearly with the phase difference between swimmers leading to multiple stable spacings at the same foil synchronization that are one wake wavelength apart \citep{Becker2015, Ramananarivo2016, Newbolt2019, Lin2019}. 

Unconstrained studies have examined the enhancement in swimming speed of foils in stable in-line formations \citep{Newbolt2019, Newbolt2022}, however, the power and efficiency benefits, or lack thereof, are still largely unknown. \cite{Heydari2021} have probed the power performance of in-line stable foils, but the phase difference between the foils is not varied, nor was leading-edge separation modeled --- a key ingredient for forming the characteristic vortex pair observed in in-line schooling that leads to branched and coherent wake modes \citep{Boschitsch2014, Kurt2018}. Additionally, considering that unconstrained foil studies show that stable formations occur when the leader and follower are thrust balanced, we can hypothesize that the power and efficiency benefits of stable formations would be small, unless they are compact, due to \textit{constrained} foil performance maps showing small power benefits for non-compact spacings when foils are thrust balanced \cite[]{Boschitsch2014,Kurt2018}. However, \textit{constrained} foil studies \citep{Kurt2021} have discovered that mismatched amplitudes between a leader and follower foil can enhance the power and efficiency performance of the collective, and \textit{unconstrained} studies \citep{Newbolt2019} have shown that in-line foils using mismatched amplitudes can also achieve stable formations. So, can \textit{unconstrained} in-line foils achieve \textit{both} high efficiency benefits and a stable formation by using mismatched amplitudes of motion?



To answer this question and provide a comprehensive investigation into the efficiency benefits of in-line schooling, new unconstrained simulations and constrained experiments of a pair of pitching foils in an in-line arrangement are conducted. Using this simple model of a school, we advance our understanding of the hydrodynamic benefits and stability of schooling formations, and their underlying mechanisms, in several ways. While using matched amplitudes of motion, we detail the mechanics of stable in-line formations, and their associated efficiency benefits, for a wide range of flow conditions defined by the Lighthill number and pitching amplitude. We show that the generation of stable formations of pitching foils relies on leading-edge separation. We derive a scaling law for the wake wavelength, $\lambda$, and demonstrate that the gap distance of stable formations and efficiency performance scales with it, rather than the nominal wake wavelength, $\lambda_0 = U/f$ (defined more below), used in previous studies \cite[]{Becker2015,Ramananarivo2016,Newbolt2019,Heydari2021,Newbolt2022}. Additionally, mismatched amplitudes of motion are discovered to substantially enhance the efficiency of stable in-line formations in both simulations and experiments. 

The paper is organized as follows. Section \ref{s:methods} describes the methodologies used throughout this study and validation of the simulations. Sections \ref{sec:separation} -- \ref{sec:scaling} present the matched amplitude simulations and development of the scaling of the wake wavelength. Lastly, Section \ref{sec:unmatched} presents the mismatched amplitude simulations and experiments.


\begin{figure}
    \centering
    \includegraphics[width=0.7\textwidth]{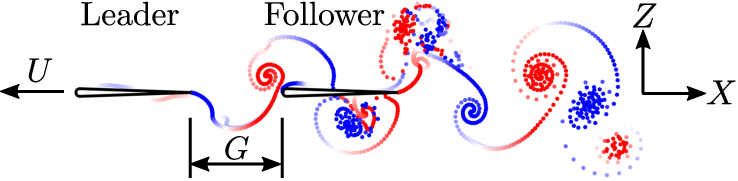}\vspace{0.in}
    \caption{Illustration of hydrofoils pitching in an in-line formation.} 
    \label{fig:foilandwake}
\end{figure}

\section{Methods} \label{s:methods}
\begin{figure}
    \centering
    \includegraphics[width=0.9\textwidth]{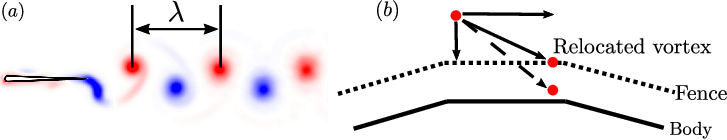}\vspace{0.in}
    \caption{$(a)$ Measurement of the wake wavelength. $(b)$ Fencing scheme and wake relocation.} 
    \label{fig:fencescheme}
\end{figure}

\subsection{Problem formulation}\label{sec:problem}
Two foils with a 7\% thick tear-drop shaped cross-section in an in-line configuration were simulated in two-dimensional potential flow by an in-house advanced boundary element method (ABEM; Figure \ref{fig:foilandwake}). Undergoing a pure pitching motion about their leading edges, a pair of leader and follower foils are free to swim in the streamwise ($X$) direction, but constrained in the cross-stream ($Z$) direction. The gap distance between the leader's trailing edge and the follower's leading edge is $G$. The properties of the leader and follower are denoted by $(\boldsymbol{\cdot})_{L}$ and $(\boldsymbol{\cdot})_{F}$, respectively. The foils' kinematics are defined by 
\begin{align} 
\theta_{L} = \theta_{0,L}\sin(2\pi ft)  \;\; \text{and} \;\; \theta_{F} = \theta_{0,F}\sin(2\pi ft-\phi),
\end{align}
where $\theta$ is the instantaneous pitching angle, $\theta_{0}$ is the pitching amplitude, $f$ is the pitching frequency, $\phi$ is the phase difference, and $t$ is time. The pitching angle is considered positive for counterclockwise pitching rotations. The dimensionless amplitude is defined as
\begin{align} 
A^* = \frac{A}{c},
\end{align}
where $A = 2 c \sin(\theta_0)$ is the peak-to-peak trailing-edge amplitude of the foil, and $c = 0.09$ \text{m} is the chord length.

The drag force, $D$, is applied to resist the motion of the foils in the streamwise direction and is used to model the viscous drag generated by the foils and from a virtual body, representing the body of a fish-like swimmer, that is not present in the computational domain \citep{Akoz2018, Ayancik2019}. Following \cite{Akoz2019} and \cite{Akoz2021}, the drag force is equal to,
\begin{align} 
D = 1/2 \,\rho u^2 Li \,c s ,
\end{align}
where $\rho$ is the fluid density, $Li$ is the Lighthill number that represents the foil's drag loading, $s = 1$ \text{m} (unit span) is the span length, and $u$ is the swimming speed of the foils. The Lighthill number varies from $0.09$ to $0.72$ in this study, which covers a range typical of aquatic animals \citep{Eloy2013,Akoz2018}. 

The thrust, $T$, is obtained from the pressure forces acting on the foils projected in the streamwise direction. The instantaneous net thrust, $T_{\text{net}} = T - D$, which is the difference between the pressure-based thrust and drag, is used to solve the streamwise equation of motion and to determine the instantaneous swimming speed (see Section \ref{sec:potential} for more details). The time-averaged swimming speed, $U$, is then defined when the cycle-averaged net thrust is zero, i.e. $\overline T_{\text{net}} = 0$. The time-averaged gap distance, $G_s$, is also averaged after the foils reach a cycle-averaged steady state. The nominal wake wavelength is defined by 
\begin{align}
\lambda_{0} = U/f,
\end{align}
which has been widely used in previous studies \cite[]{Becker2015,Ramananarivo2016,Newbolt2019,Heydari2021,Newbolt2022}. Also, the actual wake wavelength, $\lambda$, is measured as the distance between successive vortices of the same sign from the wake of the isolated leader (Figure \ref{fig:fencescheme}a), which is generally different than, though proportional to, the nominal wavelength. The Strouhal number and reduced frequency are defined by
\begin{align}
St = \frac{fA}{U} \;\; \text{and} \;\; k = \frac{fc}{U},
\end{align}
respectively. 

The power input to the fluid, $P$, is obtained by integrating the inner product between the pressure force vectors and local kinematic velocity vectors along the surface of the foil (defined precisely in Section \ref{sec:potential}), which is equivalent to the product of the pitching moment about the pitching axis and the angular velocity \cite[]{moored2018inviscid}. The time-averaged thrust and power coefficients are then defined by,
\begin{align}
C_{T} = 
\frac{\overline{T}}{1/2\, \rho U^2 c s}\;\; \text{and} \;\; C_{P}=\frac{\overline{P}}{1/2\, \rho U^3 c s},
\end{align}
respectively. The collective and isolated efficiency are calculated as
\begin{align} 
\eta_{c} = \frac{(\overline{T}_{L}+\overline{T}_{F})U}{\overline{P}_{L}+\overline{P}_{F}} \;\; \text{and} \;\;\eta_{iso} = \frac{\overline{T}_{L,\text{iso}}U_{L,\text{iso}}+
\overline{T}_{F,\text{iso}}U_{F,\text{iso}}}{\overline{P}_{L,\text{iso}}+\overline{P}_{F,\text{iso}}}, \label{eq:eff}
\end{align}
respectively. Furthermore, the collective efficiency is normalized by the isolated efficiency to give
\begin{align} 
\eta_c^* = \frac{\eta_{c}}{\eta_{iso}}.  
\end{align}
When the normalized collective efficiency is greater than one, the two schooling foils are more efficient swimming together than in isolation and \textit{vice versa}. 

\subsection{Advanced boundary element method}
Our in-house potential flow solver is an advanced boundary element method that accounts for leading-edge flow separation of the hydrofoils and employs a fencing scheme to prevent wake vortices from penetrating the hydrofoils.

\subsubsection{Governing equations}\label{sec:potential}
For potential flow that is incompressible, irrotational, and inviscid, Laplace's equation serves as the governing equation,
\begin{align} 
\nabla^2 \Phi^* \, = \,0.
\end{align}
Here $\Phi^*$ is the perturbation potential in an inertial reference frame fixed to the undisturbed fluid. Additionally, the no-flux boundary condition is satisfied on the two foils' body surfaces, $S_{b_{L}}$ and $S_{b_{F}}$,
\begin{align} 
 \bold{n} \boldsymbol{\cdot} \nabla \Phi^* \, = \, \bold{n} \boldsymbol{\cdot} (\bold{u_{rel}} + \bold{u}), 
\end{align}
where $\bold{u_{rel}}$ is the  relative velocity to the body-fixed reference frame fixed at a foil's leading edge, and $\bold{n}$ is the normal vector pointing outward from the body surface. Additionally, the far-field condition that the flow disturbances caused by the body must decay far away must be satisfied,
\begin{align} 
 \underset{|r|\rightarrow\infty}{\text{lim}}( \nabla \Phi^*)\,=\,0, 
\end{align}
where $|r|$ is the distance from a target point to the body.

Following \cite{lowspeed2001} and \citet{Moored2018}, $\Phi^*$ can be solved by the boundary integral equation that considers a combination of constant-strength sources, $\sigma$, and doublets, $\mu$, on the body surface and constant-strength doublets, $\mu_{w}$, on the two foils' wake surfaces,
\begin{gather} 
 \Phi^*_{i}(r)  =  \iint_{S_b^*} [\sigma(r_{0})G(r;r_{0}) - \mu(r_{0})\bold{n}\boldsymbol{\cdot}\nabla G(r;r_{0})]  \text{d} S - \iint_{S_w^*} \mu_{w}(r_{0})\bold{n} \boldsymbol{\cdot} \nabla G(r;r_{0}) \text{d} S \label{eq:green}
 \end{gather}
 with
 \begin{gather}
\sigma = \bold{n} \boldsymbol{\cdot} \nabla (\Phi^* - \Phi^*_{i}) \;\;\text{and} \;\; -\!\mu = \Phi^* - \Phi^*_{i},
\end{gather}
where $\Phi^*_{i}$ is the internal potential, $S_b^* = S_{b,L} + S_{b,F}$ is the combined body surfaces, $S_w^* = S_{w,L} + S_{w,F}$ is the combined wake surfaces, $r_{0}$ is a source or doublet's location, $r$ is the location of the target point, and the Green's function is $G(r;r_{0})=(1/2\pi) \ln | r-r_{0}|$. The Dirichlet boundary condition is enforced by setting the internal potential to be zero, $\Phi^*_{i}$ = 0, which gives
\begin{gather} 
\sigma = \bold{n} \boldsymbol{\cdot} \nabla \Phi^* = \bold{n} \boldsymbol{\cdot} (\bold{u_{rel}} + \bold{u}) \;\; \text{and} \;\; -\!\mu = \Phi^*. \label{sigmaandmu}
\end{gather}
Consequently, $\sigma$ becomes a known value, and $\Phi^*$ can be calculated by solving $\mu$ over the body. Conveniently, the far-field condition is implicitly satisfied by the source and doublet elements.

Each of the foils' body surfaces are discretized into $N$ panels. The doublet strength of every body panel is unknown. Also, the wake surfaces of the foils are discretized into doublet panels. At each time step, a trailing edge panel will shed from each foil into the fluid, whose strength won't change in time and a new trailing edge panel will form. The strength of the trailing-edge panel is determined by enforcing the Kutta condition, that is, zero vorticity at the trailing edge,
\begin{align} 
\mu_{TE} = \mu_{+} - \mu_{-},
\end{align}
where $\mu_{TE}$ is the strength of the trailing-edge panel, $\mu_{+}$ is the doublet strength of the body panel on the upper surface at the trailing edge, and $\mu_{-}$ is the doublet strength of the body panel on the lower surface at the trailing edge. The trailing-edge panel is oriented along a line that bisects the upper and lower surfaces of the body at the trailing edge, whose length is set to $0.4 U \Delta t$, where the simulation time step is $\Delta t$. There are $2N$ unknowns $\mu$, and they can be solved by $2N$ equations (\ref{eq:green}) enforced at $2N$ collocation points. These points are located at the center of panels, but moved inside the body along the surface normal vector by $10\%$ of the local body thickness. 

Then the pressure field over the body, $p$, can be calculated by the unsteady Bernoulli equation,
\begin{align} \label{eq:unsteadybernoulli} 
p(x,z,t)=-\rho \frac{\partial \Phi^*}{\partial t} + \rho (\mathbf{u_{rel}} + \mathbf{u}) \boldsymbol{\cdot} \nabla \Phi^* -\rho \frac{(\nabla \Phi^*)^2}{2}.
\end{align}

The thrust, $T$, and lift, $L$, can be calculated by integrating the normal forces from the pressure fields, projected in the $x$ and $z$ directions, acting on the body surfaces, 
\begin{align} 
T = \int_{S_{b}} - p  \bold{n} \boldsymbol{\cdot} \bold{x} \, \text{d} S \;\; \text{and} \;\;  L = \int_{S_{b}} - p  \bold{n} \boldsymbol{\cdot} \bold{z} \, \text{d} S,
\end{align}
where $\bold{x}$ is the unit vector in the streamwise direction, and $\bold{z}$ is the unit vector in the cross-stream direction. Power consumption is calculated by
\begin{align}
P = \int_{S_{b}}  - p  \bold{n} \boldsymbol{\cdot} \bold{u_{rel}} \, \text{d} S.
\end{align}

To calculate the foils' free-swimming motion in the streamwise direction, the velocity and location of the foil at the $(n + 1)^\text{th}$ time step are calculated by forward differencing \citep{Moored2018, Ayancik2020},
\begin{align} 
u^{n+1} = u^{n} + \frac{T_{net}^{n}}{m} \Delta t \;\; \text{and} \;\; X^{n+1} = X^{n} + \frac{1}{2}(u^{n+1}+u^{n}) \Delta t,
\end{align}
where $X$ is the location of a foil's leading edge in the streamwise direction, and $m$ is the mass of the foil, which is fixed in this study at three times the characteristic added mass of the foils, i.e. $m = 3\rho s c^2$.

The number of body panels per foil and time steps per pitching cycle were both set to $250$ based on convergence studies, which showed that stable formations change by less than $2\%$ when both are doubled from $250$ to $500$, independently.

\subsubsection{Leading-edge flow separation}
Following \cite{Ramesh2014}, a leading-edge separation mechanism is coupled with the potential flow solver based on the leading-edge suction parameter (LESP), which indicates the suction strength at a foil's leading edge. Based on \cite{Narsipur2020}, the instantaneous LESP value is calculated by
\begin{align} \label{eq:scal1} 
\text{LESP}(t) = A_{0}(t) = -\frac{1}{\pi u_{\text{net}}}\int_{0}^{\pi}\frac{\partial \Phi^*_{B}}{\partial z} \text{d} \alpha
\end{align}
with
\begin{align}
x = \frac{c}{2}(1-\cos{\alpha})
\end{align}
and
\begin{align}
u_{\text{net}}(t) = \sqrt{\left[u(t) + \frac{1}{2}\dot{\theta}(t)c\sin{\theta(t)}\right]^2 + \left[\frac{1}{2}\dot{\theta}(t)c\cos{\theta(t)}\right]},
\end{align}
where $\Phi^*_{B}$ is the perturbation potential generated by bound vortices, $\alpha$ is a transformation variable associated with the chordwise coordinate $x$, and $z$ is the direction normal to a foil's chord line. According to \cite{Ramesh2014}, a leading-edge vortex will be shed when $|\text{LESP}|$ exceeds its critical value $\text{LESP}_{\text{crit}}$, and, by shedding a leading edge vortex element, $|\text{LESP}|$ will be brought back to its critical value,
\begin{align}\label{eq:crilesp}
|\text{LESP}(t)| =  \text{LESP}_{\text{crit}}.
\end{align}
Additionally, $\text{LESP}_\text{crit}$ depends on the foil's shape and Reynolds number. 

At each time step, $2N$ equations (\ref{eq:green}) are solved without shedding any leading-edge vortices first, then $\text{LESP}$ is calculated. If the $|\text{LESP}|$ of the foils does not exceed their critical values then the simulation moves to the next time step. If the $|\text{LESP}|$ of the foils exceed their critical values, an additional equation~(\ref{eq:crilesp}) will be formulated for and an additional unknown leading-edge vortex will be shed from each foil exceeding its critical value. 

\subsubsection{Wake-body interaction}
To represent the advection of vorticity with the local flow velocity a so-called ``wake roll-up" or wake element advection scheme must be applied at every time step. To calculate the induced flow velocity from the wake elements on themselves, the elements must be desingularised for numerical stability. Following previous work \cite[]{Vatistas1991,Ramesh2014,Sedky2020}, the induced velocity of a constant strength doublet element is equivalent to two counter-rotating point vortices located at the element end points, which can be calculated using the desingularised Lamb-Oseen two-dimensional vortex model, 
\begin{align}\label{eq:lamboseen}
\boldsymbol{u}_{i}(\boldsymbol{r}) = \frac{\Gamma_{i}}{2\pi}\frac{{\bold{y}\times (\boldsymbol{r}-\boldsymbol{r_{i}})}}{\sqrt{|\boldsymbol{r}-\boldsymbol{r_{i}}|^4 + \delta^4}} \;\; \text{and} \;\; \boldsymbol{r}-\boldsymbol{r_{i}} = (x-x_{i})\boldsymbol{x}+ (z-z_{i})\boldsymbol{z}.
\end{align}
Here $(x,z)$ is the coordinate of the influence point, $(x_{i},z_{i})$ is the coordinate of the $i^\text{th}$ vortex, $\Gamma_{i}$ is the strength of the $i^\text{th}$ vortex, and $\delta$ is the  desingularisation parameter. The desingularisation parameter is a free parameter and is used to mimic the effect of viscosity in real fluids by giving each point vortex a core radius. We choose the lowest value for $\delta$ that prevents numerical instability during wake advection. It increases from $\delta = 0.06$ to $0.25$ when the pitching amplitude increases from $\theta_0 = 7.5^{\circ}$ to $30^{\circ}$.

In order to prevent wake vortex elements from penetrating the foils during advection, a fence with thickness of 3\% of the chord length is built around each foil. When a vortex penetrates the body a fencing scheme is applied and it is relocated such that the displacement of the element end point normal to the surface is cancelled while the displacement tangent to the surface is preserved (Figure \ref{fig:fencescheme}b).

\subsection{Experimental setup}

\begin{figure}
    \centering
    \includegraphics[width=\textwidth]{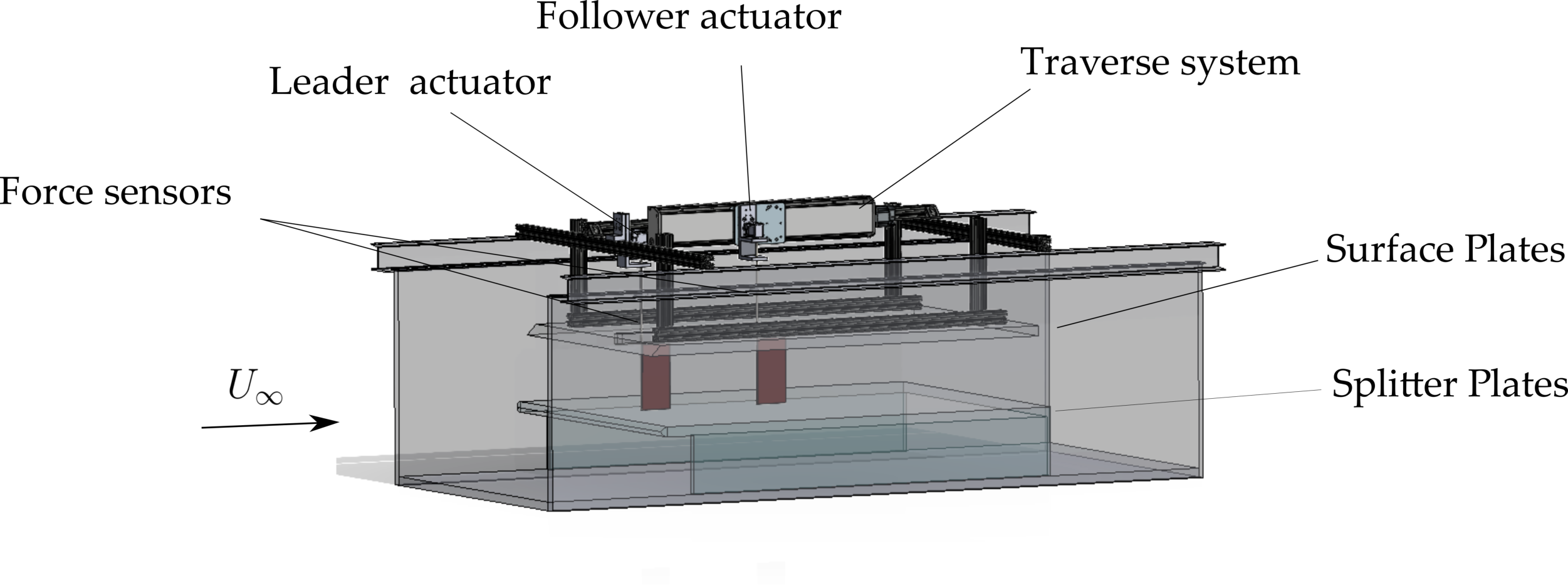}
    \caption{Schematic of experimental setup.}
    \label{fig:expschem}
\end{figure}
To identify potential efficiency benefits for unconstrained foils and to verify the mismatched amplitude simulation results, we conducted experiments on two constrained foils in in-line arrangements undergoing the same kinematics as the unconstrained foils. Measurements occurred in a closed-loop water tunnel (Figure~\ref{fig:expschem}) with a splitter and surface plate used to create a nominally two-dimensional flow. Two identical hydrofoils, designated as the leader and follower, were used and had a rectangular planform shape, a chord length of $c = 0.09$ m, a span length of $s = 0.18$ m, giving them an aspect ratio of two. The foils were made of acrylonitrile butadiene styrene (ABS) and featured the same cross-sectional shape as foils in the simulations. The pitching kinematics of the foils were controlled by Dynamixel MX-64T servomotors and the pitching axis was set at the leading edge of the foils. Details of the kinematics of the foils can be found in Section \ref{sec:problem}. The phase difference of the foils varies between $0\leq \phi \leq 2\pi$  with increments of $\pi/4$, producing eight different phase synchronies for each arrangement. The gap distance was varied between $0.25c \leq G \leq 1.75c$ with increments of $0.25c$ by utilizing an automated traverse mechanism. The pitching frequency and amplitude, phase difference, and flow speed of the constrained experiments were kept consistent with those of the unconstrained simulations. Two ATI Mini-40E six-axis force sensors were used on both leader and follower foils to measure the thrust and pitching moment acting on each foil. Further details of the force measurement setup can be found in \cite{Kurt2018} and \cite{Han2022}. During each trial the thrust and power were time-averaged over $60$ pitching cycles and these mean values are averaged over five trials and reported. Collective and isolated efficiencies are calculated following equation (\ref{eq:eff}). The maximum standard deviation of the five trials for the measured efficiencies is $2.5\%$.

\subsection{Numerical validation}
In order to validate the in-house ABEM solver three validation cases are presented, which cover a foil performing a pitch maneuver, a foil in combined heaving and pitching motion, and two foils in an in-line schooling formation. 

\subsubsection{Pitch maneuver}

\begin{figure}
    \centering
    \includegraphics[width=\textwidth]{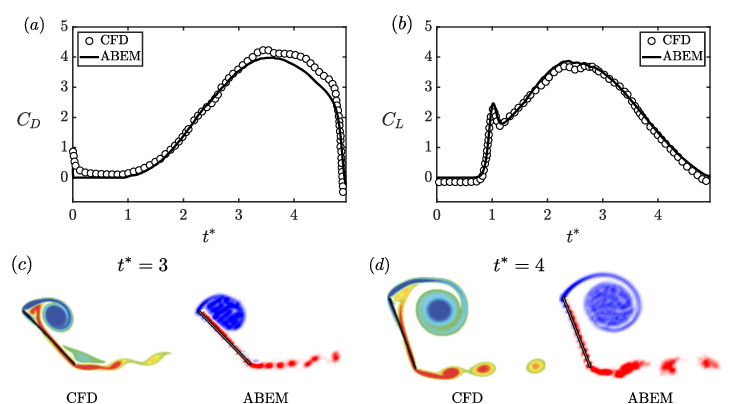}\vspace{0.in}
    \vspace{-0.1in}
    \caption{ABEM potential flow solver shows good agreement with CFD results calculated using a viscous vortex particle method \cite[]{Wang2013} for a pitch maneuver. The forces are compared in $(a)$ for the drag coefficient and in $(b)$ for the lift coefficient. The flow structures are compared in $(c)$ and $(d)$ at different times.} 
   \label{fig:validationpitchma}
\end{figure}

Figure \ref{fig:validationpitchma} present the forces and flow structures simulated with the ABEM solver compared to previous CFD results from a viscous vortex particle method \cite[]{Wang2013}. A $2.3\%$ thick flat plate pitching about its leading edge was simulated at $ Re = 1000$. The critical leading-edge suction parameter was chosen to be $\text{LESP}_\text{crit} = 0.1$ \cite[]{Ramesh2014}. The instantaneous pitch angle of the foil was given by
\begin{align}
\theta(t^*) = \theta_{0} \frac{G}{G_{\text{max}}},
\end{align}
with
\begin{align}
G(t^*) = \log \left\{ \frac{\cosh{\left[a_s(t^*-t_{1}^*)\right]}}{\cosh{\left[a_s(t^*-t_{2}^*)\right]}} \right\} - a_s(t_{1}^*-t_{2}^*),
\end{align}
where $a_s = 11$ is the smoothing parameter,  $t^*$ is the dimensionless time with $t^* = tU_{\infty}/c$ and $U_{\infty}/c = 1$, $t_{1}^*$ and $t_{2}^*$ are two constants with $t_{1}^* = 1$ and $t_{2}^* = 1 + 5\pi/4$, and $G_{\text{max}}$ is the maximum of $G$ when $t^*$ varies from $0$ to $5$. In Figure \ref{fig:validationpitchma}, it is found that both force data and flow structures are in good agreement with the CFD results.

\subsubsection{Combined heaving and pitching}
\begin{figure}
    \centering
    \includegraphics[width=\textwidth]{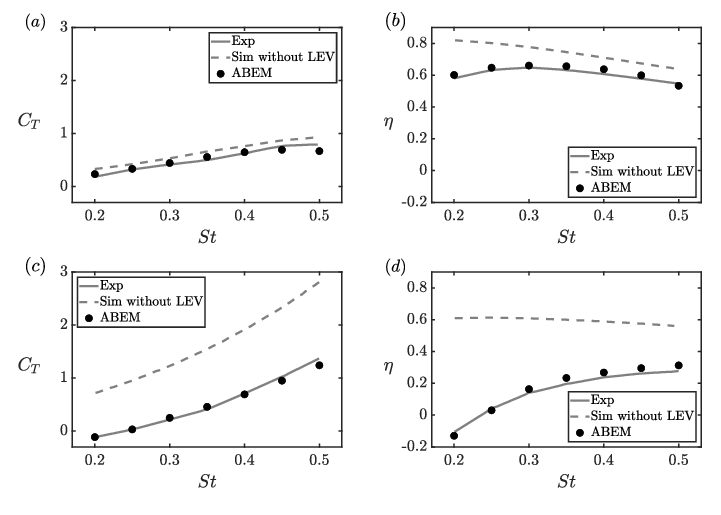}\vspace{0.in}
    \vspace{-0.1in}
    \caption{Validation of a hydrofoil performing combined heaving and pitching motion. The $(a)$ thrust coefficient and $(b)$ efficiency at $\alpha_{\text{max}}=15^{\circ}$. The $(c)$ thrust coefficient and $(d)$ efficiency at $\alpha_{\text{max}}=40^{\circ}$. ABEM simulation data with and without a leading-edge separation model are represented by black circles and dashed lines, respectively. Experimental data from \cite{Read2003} is represented by solid lines.}
   \label{fig:validationcombined}
\end{figure}

Figure \ref{fig:validationcombined} presents ABEM simulation data compared against previous experiments \cite[]{Read2003} of a NACA $0012$ hydrofoil performing combined heaving and pitching motion at $Re = 40,000$. The foil's kinematics are given as,
\begin{align}
h(t) = h_{0}\sin{(2\pi f t)} \;\; \text{and} \;\; \theta(t) = \theta_{0}\sin{(2\pi f t + \pi/2)},
\end{align}
where $h(t)$ is the foil's heaving motion, and the foil's heaving amplitude, $h_{0}$, is fixed at one chord length with $c = 0.1\,\text{m}$. The instantaneous angle of attack of the foil is calculated by
\begin{align}
\alpha(t) = \theta(t) - \tan^{-1} \left[\frac{\dot{h}(t)}{U_{\infty}}\right].
\end{align}
The free-stream velocity is $U_\infty = 0.4\,\text{m/s}$. The maximum angle of attack, $\alpha_{\text{max}}$, is the maximum $\alpha$ within one cycle of oscillation, and simulations were conducted at two maximum angles of attack that were $\alpha_\text{max} = 15^{\circ}$ and $40^{\circ}$. \cite{Narsipur2020} showed that $\text{LESP}_{\text{crit}}$ varies when the non-dimensional rate of change of angle of attack, $K = \dot{\alpha}c/2U_{\infty}$, has a large variation, which happens in the combined pitching and heaving motion. According to \cite{Narsipur2020},  $\text{LESP}_{\text{crit}}$ increases from $0.15$ to $0.2$ based on the average rate of change of angle of attack, $K_{\text{avg}} = 2\alpha_{\text{max}}cf/U_{\infty}$. Figure \ref{fig:validationcombined} presents the comparison of thrust and efficiency at the two maximum angles of attack between the ABEM simulations and experiments. It shows that the ABEM simulations are in good agreement with the experimental data. In addition, it is discovered that thrust and efficiency are significantly over-predicted when there is no leading-edge separation model and $\alpha_{\text{max}}=40^{\circ}$. 

\subsubsection{In-line schooling of a pair of pitching foils}
\begin{figure}
    \centering
    \includegraphics[width=\textwidth]{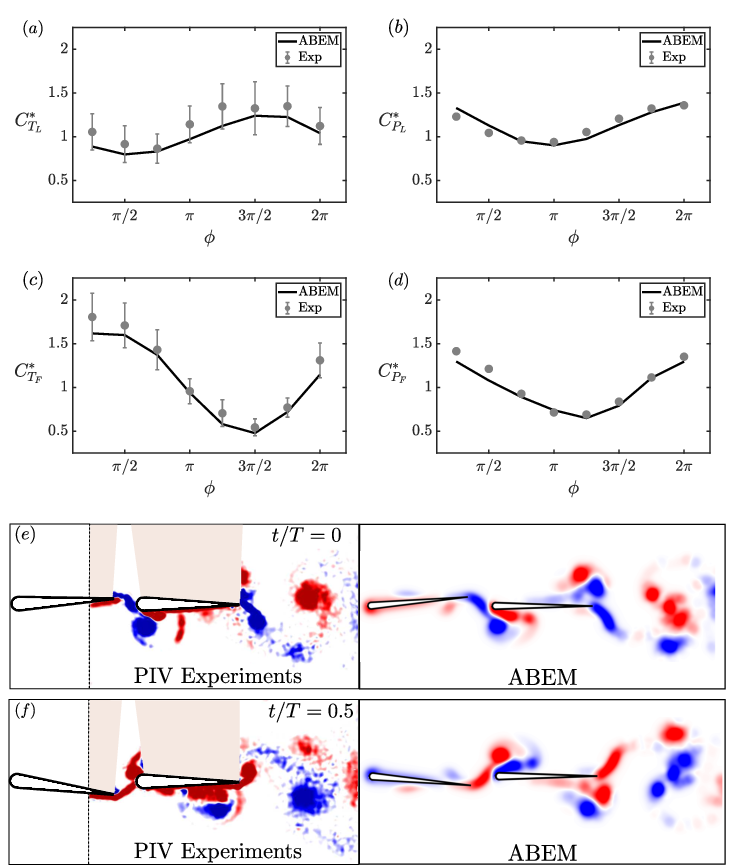}\vspace{0.in}
    \vspace{-0.1in}
    \caption{ Validation of in-line schooling simulations against previous experimental data \cite[]{Kurt2018}. Normalized thrust coefficient for $(a)$ the leader foil and $(c)$ the follower foil. Normalized power coefficient for $(b)$ the leader foil and $(d)$ the follower foil. Error bars represent $\pm$ one standard deviation. Comparison of vorticity between the experiments (left) and simulations (right) for $\phi = \pi/4$ at $(e)$ $t/T = 0$ and $(f)$ $t/T = 0.5$.}
   \label{fig:validationschool}
\end{figure}
Figure \ref{fig:validationschool} presents simulations of constrained pitching foils in an in-line schooling formation.  Simulations were conducted for two pitching foils with a 7\% thick teardrop cross-sectional shape and validated against previous experimental data \cite[]{Kurt2018}. The gap distance is fixed at $G/c=0.25$, and the phase difference varies from $0 \leq \phi \leq 7\pi/4$. The foils' pitching frequency and amplitude are fixed at $f = 0.75$ Hz and $\theta_0 = 7.5^{\circ}$, respectively. In addition, the foils' chord length is fixed at $c = 0.09$ m, and the free-stream velocity is fixed at $U_\infty = 0.071$ m/s. According to \cite{Ramesh2014}, $\text{LESP}_\text{crit} = 0.1$ was determined by finding  the value that can provide the best match with the experimental thrust data of an isolated foil. The normalized time-averaged thrust and power coefficients of the leader are defined by
\begin{align}
C_{T_L}^* = \frac{\overline{C}_{T_{L}}}{\overline{C}_{T_{\text{iso}}}} \;\; \text{and} \;\; C_{P_L}^* = \frac{\overline{C}_{P_{L}}}{\overline{C}_{P_{\text{iso}}}},
\end{align}
respectively, where $\overline{C}_{T_{\text{iso}}}$ and $\overline{C}_{P_{\text{iso}}}$ are the time-averaged thrust and power coefficients of an isolated foil, respectively. Similarly, the normalized time-averaged thrust and power coefficients of the follower can be defined.
Figure \ref{fig:validationschool}a--\ref{fig:validationschool}d shows a good agreement in the normalized thrust and power coefficients between ABEM simulations and experiments across the full range of phase difference. Figure \ref{fig:validationschool}e and \ref{fig:validationschool}f highlight that the ABEM simulations also capture the main vortex structures observed in the PIV data further validating the simulations.
\begin{table}
\centering
\begin{tabular}{ccccccccccccccc}
  &  &  Varying $\text{LESP}_{\text{crit}}$ & &  Varying  $Li$ & &  Varying $\theta_{0}$ & &  Varying $\theta_{0,L}/\theta_{0,F}$ \\ 
  $\phi$ & & $0-2\pi$ & &   $0-2\pi$ & &   $0-2\pi$ & &  $0-2\pi$\\
$\text{LESP}_{\text{crit}}$ & & $0.1$, $100$ & &   $0.1$ & &   $0.1$ & &  $0.1$\\
$Li$ & & $0.09-0.72$ & &   $0.09-0.72$ & &   $0.36$ & &  $0.18$, $0.36$\\
$\theta_{0}$ & & $7.5^{\circ}-30^{\circ}$ & &  $7.5^{\circ}$ &  &$7.5^{\circ}-30^{\circ}$& &  $7.5^{\circ}$, $15^{\circ}$\\
$\theta_{0,L}/\theta_{0}$ & & $1$ & &   $1$ & & $1$& &  $1.2$\\
$\theta_{0,F}/\theta_{0}$ & & $1$ & &   $1$ & & $1$& &  $1$\\
\end{tabular}
\caption{ \label{table:parameter} Input parameters and variables used in the current study.}
\end{table}

\section{Results}
Sections \ref{sec:separation}--\ref{sec:amp} present results from simulations of two streamwise unconstrained pitching hydrofoils swimming in an in-line formation over a wide range of Lighthill number, \textit{matched} pitching amplitude, and phase difference. In addition, simulations with and without leading-edge flow separation were conducted to probe its effect on the formation of in-line stable arrangements. Suppressing leading-edge flow separation was achieved by giving $\text{LESP}_{\text{crit}}$ an effectively infinite value (chosen to be 100 in this study).  The input variables used in these sections are detailed in Table \ref{table:parameter}. Section \ref{sec:unmatched} presents constrained experiments and unconstrained simulations of \textit{mismatched} amplitudes where the leader has a larger amplitude than the follower. The input variables for this section are also detailed in Table \ref{table:parameter}.

\subsection{Varying the condition of leading-edge flow separation}\label{sec:separation}

\begin{figure}
    \centering
    \includegraphics[width=\textwidth]{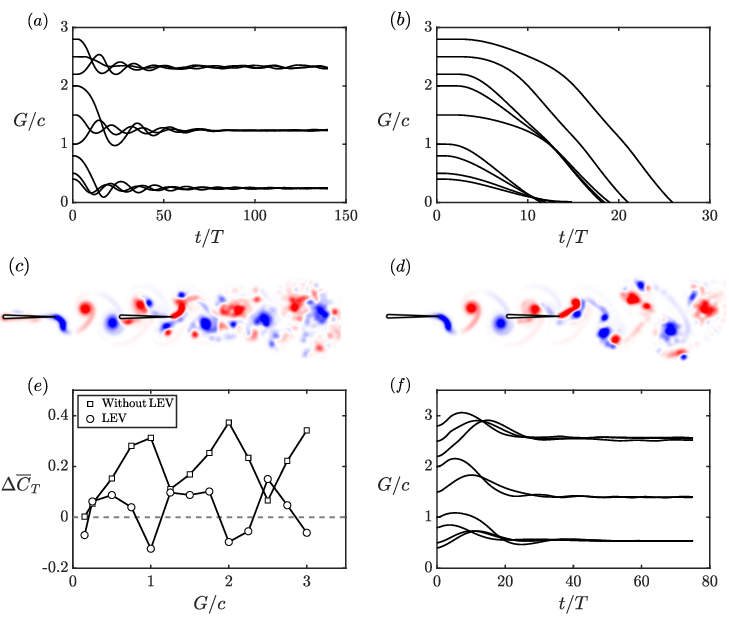}\vspace{0.in}
    \vspace{-0.1in}
    \caption{Leading-edge flow separation is critical for generating stable in-line formations. Trajectories for $Li = 0.36$, $\theta_{0} = 7.5^{\circ}$ and $\phi=\pi$: $(a)$ simulations with leading-edge flow separation, $\text{LESP}_{\text{crit}}=0.1$, and $(b)$ simulations without leading-edge flow separation, $\text{LESP}_{\text{crit}}=100$. Vorticity for a simulation $(c)$ with and $(d)$ without leading-edge flow separation when $G/c= 1.24$. $(e)$ Thrust coefficient difference ($\Delta\overline{C}_{T}=\overline{C}_{T,F}-\overline{C}_{T,L}$) from constrained simulations with and without leading-edge separation. $(f)$ Time-varying gap distance from simulations without leading-edge separation, but  when the follower has a larger Lighthill number than the leader of $Li = 0.54$.}
   \label{fig:trajectory}
\end{figure}

A representative simulation case with and without leading-edge flow separation is presented in Figure \ref{fig:trajectory}, where $Li = 0.36$, $ \theta_{0} = 7.5^{\circ}$, and $\phi=\pi$. Figure \ref{fig:trajectory}a shows that the leader and follower foils' wide range of initial gap distances evolve to settle into three distinct stable gap distances when leading-edge flow separation is present, as observed previously in experiments \cite[]{Becker2015, Ramananarivo2016, Newbolt2019}. However, Figure \ref{fig:trajectory}b shows that when leading-edge separation is \textit{suppressed} the follower always collides into the leader ($G/c \rightarrow 0 $) regardless of its initial gap distance. Furthermore, this phenomenon occurs across the whole range of Lighthill number, pitching amplitude, and phase difference shown in Table \ref{table:parameter} (varying $\text{LESP}_{\text{crit}}$). Figure \ref{fig:trajectory}c and \ref{fig:trajectory}d show that the characteristic vortex pairs that form on the top and bottom of the follower experiencing in-line interactions \cite[]{Boschitsch2014,Kurt2018} disappear when separation is suppressed. 

To better understand the role that leading-edge separation is playing to create stable formations, we conducted additional constrained simulations at the free-swimming velocity of an isolated leader, which is within $5\%$ of the free-swimming velocity for the leader-follower pairs in Figure \ref{fig:trajectory}a. The constrained simulations allow us to map out the thrust difference between the leader and follower with varying gap distance, which is presented in Figure \ref{fig:trajectory}e. For both cases of with and without leading-edge separation, the thrust coefficient difference shows a sinusoidal-like variation with the gap distance \cite[]{Boschitsch2014,baddoo2023generalization}. For the case with leading-edge flow separation, the thrust curve is shifted lower than the case without separation and it crosses the zero thrust difference line indicating locations for in-line equilibrium formations, however, only the zero crossings with a positive slope represent \textit{stable} formations \cite[]{Ramananarivo2016}. As shown in \cite{Boschitsch2014}, the synchronization of the follower's motion to the impinging vortex shed from the leader can either enhance thrust or induce drag on the follower, making the difference in thrust coefficients fluctuate about zero. Also, in \cite{baddoo2023generalization} it was shown that without leading-edge separation pitching in-line foils never reached stable formations, that is, the follower always produced more thrust than the leader, which is corroborated in our simulations (Figure \ref{fig:trajectory}b). The key idea is that leading-edge separation is seen to act as an \textit{additional dynamic drag source} on the follower whose magnitude varies depending upon the phasing between the impinging vortex and the follower's pitching motion. This is evident in the ``inversion" of the thrust difference curve where the peaks in thrust difference when separation is suppressed (with a net thrust for the follower) become troughs in thrust difference when separation is present (with a net drag for the follower). We hypothesize that leading-edge separation acts as a dynamic drag source on the follower, where a constant additional drag source 
may provide qualitatively similar results such as creating stable formations, but likely quantitatively different dynamics. 

To verify this idea, new free-swimming simulations are presented in Figure \ref{fig:trajectory}f where there is no leading-edge separation present, but the follower's Lighthill number is increased to  $Li = 0.54$ while the leader's remains at $Li = 0.36$. This increase in Lighthill number of the follower increases the drag of the follower in a way that is independent of the phase of motion relative to the impinging vortex wake.
As expected, Figure \ref{fig:trajectory}f reveals that adding an additional \textit{constant} drag source to the follower has the effect of creating in-line stable formations as does leading-edge separation. However, the dynamic response of the follower and the final stable gap distances are observed to be quantitatively different for a constant drag source than for the dynamic drag source incurred by leading-edge separation. This verifies that for pitching foils, an increased drag on the follower is critical in creating in-line stable formations and to capture the dynamic response and stable gap distances accurately leading-edge separation must be modeled, that is, a constant additional drag on the follower does not suffice. 

\subsection{Varying the Lighthill number}\label{sec:Li}
\begin{figure}
    \centering
    \includegraphics[width=\textwidth]{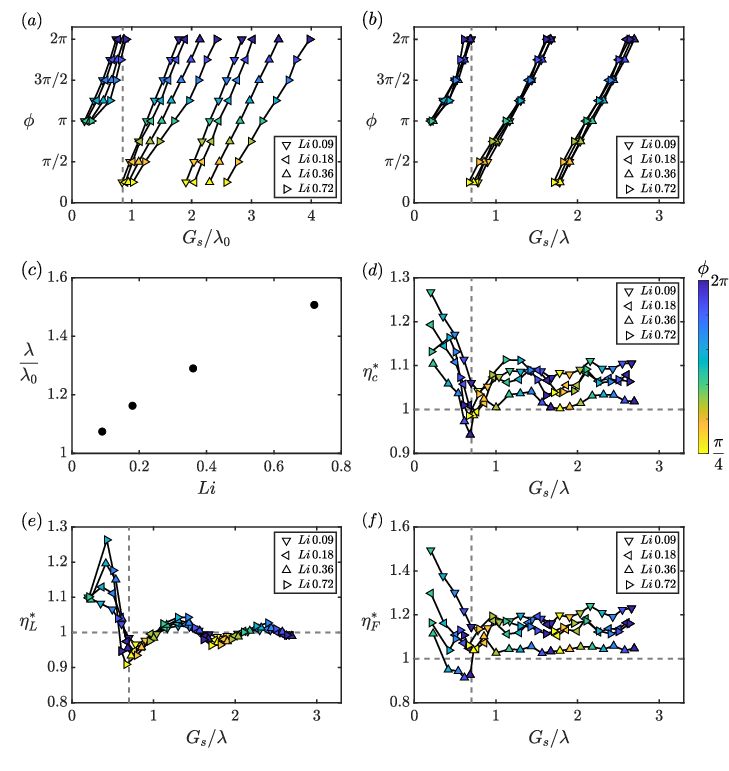}\vspace{0.in}
    \vspace{-0.1in}
    \caption{In-line stable gap distances scale with the actual wake wavelength, $\lambda$, rather than the nominal wavelength, $\lambda_{0}$, for varying $Li$. The pitching amplitude is $\theta_0 = 7.5^{\circ}$. The gray vertical dashed lines represent $G_{s}/\lambda_{0}=0.85$ and $G_{s}/\lambda=0.7$ where the wavelengths are calculated/measured for the isolated leader. Stable gap distances scaled by $(a)$ $\lambda_{0}$ and $(b)$ $\lambda$. $(c)$ Wake wavelength compared to the nominal wavelength as a function of the Lighthill number. $(d)$ Normalized collective efficiency as a function of the normalized gap distance. $(e)$ Normalized leader efficiency as a function of the normalized gap distance. $(f)$ Normalized follower efficiency as a function of the normalized gap distance.}
   \label{fig:varyli}
\end{figure}

For the free-swimming simulations presented in this section, the pitching amplitude of the foils is fixed at $\theta_0 = 7.5^{\circ}$ while the Lighthill number and phase synchrony vary over the ranges of $0.09 \leq Li \leq 0.72$ and $0 \leq \phi \leq 2\pi$, respectively (see varying $Li$ in Table \ref{table:parameter}). Figure \ref{fig:varyli}a presents the stable gap distances of the foils normalized by the nominal wake wavelength as a function of the phase difference and Lighthill number. When the stable position is normalized by $\lambda_{0}$, used in previous studies \citep{Becker2015,Ramananarivo2016,Newbolt2019,Heydari2021,Newbolt2022}, there is no collapse of the data with varying $Li$. In contrast, Figure \ref{fig:varyli}b reveals that a good collapse of the data occurs when the stable gap distances are normalized by the actual wake wavelength of an isolated foil, $\lambda$. Figure \ref{fig:varyli}a and \ref{fig:varyli}b both exhibit linear relationships between the phase difference and the normalized stable position. But for $G_{s}/\lambda < 0.7$ and $G_{s}/\lambda_{0} < 0.85$, the relationship between the phase difference and the normalized stable position is observed to have a slightly non-linear trend  when the Lighthill number is increased from $Li = 0.09$ to $0.72$. Figure \ref{fig:varyli}c reveals that the ratio of $\lambda/\lambda_{0}$ increases with increasing $Li$, which leads to the difference between $\lambda$ and $\lambda_{0}$ with respect to scaling the stable positions. 

Figure \ref{fig:varyli}d presents the normalized collective efficiency as a function of the normalized stable gap distances. There is an efficiency benefit for schooling ($\eta_{c}^* > 1$) at nearly all stable positions and the maximum efficiency enhancement is achieved for compact arrangements with $G_{s}/\lambda<0.7$ for all $Li$. Regardless of the Lighthill number, the collective efficiency has successive peaks that have gap distances one wavelength apart and nearly all of the peaks in efficiency occur for $\phi \approx \pi$ while nearly all of the troughs occur for $\phi \approx 0$. Moreover, the maximum efficiency enhancement decreases with increasing Lighthill number from $Li = 0.09$ to $0.36$ and then increases with further increases in $Li$. Even though the data represent the same impingement timing between the leader's vortices and the follower's motion due to the proportional change in spacing and phase synchrony, in non-compact formations the collective efficiency enhancement differs. This indicates that more than a vortex-body mechanism is at play. In fact, there are both efficiency enhancements on the leader (Figure \ref{fig:varyli}e) from body-body interactions and on the follower (Figure \ref{fig:varyli}f) from vortex-body interactions. The leader's efficiency presented in Figure \ref{fig:varyli}e shows that a phase synchrony of $\phi \approx \pi$ will maximize efficiency while $\phi \approx 0$ will minimize it regardless of the gap spacing. This is an indication of a direct body-to-body interaction since the effect is upstream and only depends upon the phase synchrony between the foils. In contrast, the follower's efficiency presented in Figure \ref{fig:varyli}f shows that when the gap spacing and phase synchrony change proportionally then there is no preferred phase to maximize the efficiency, at least in non-compact formations with $G_{s}/\lambda > 0.7$. This is indicative of a vortex-body interaction.


\subsection{Varying the pitching amplitude}\label{sec:amp}
\begin{figure}
    \centering
    \includegraphics[width=\textwidth]{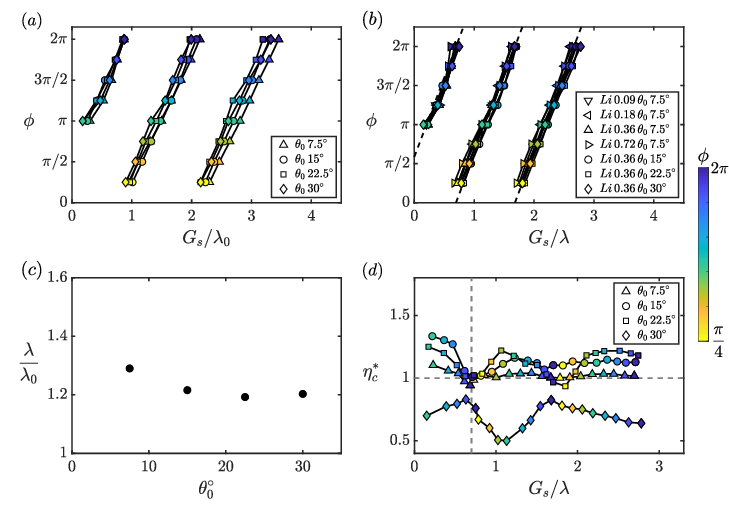}\vspace{0.in}
    \vspace{-0.1in}
    \caption{The stable gap distances scale with the leader's wake wavelength regardless of variations in $Li$ or $\theta_{0}$. The Lighthill number is $Li = 0.36$ for $(a)$, $(c)$, and $(d)$. The wavelengths are calculated/measured for the isolated leader. Stable gap distances scaled by $(a)$ $\lambda_{0}$ and $(b)$ $\lambda$ (with the data of varying $Li$ included). Black dashed lines represent the linear relationship $G_s/\lambda = \phi/(2\pi) -0.29$. $(c)$ Wake wavelength compared to the nominal wavelength as a function of the pitching amplitude. $(d)$ Normalized collective efficiency as a function of the normalized gap distance. The gray vertical dashed line represents $G_s/\lambda=0.7$.}
   \label{fig:varyamp}
\end{figure}

For the free-swimming simulations presented in this section, the Lighthill number is fixed at $Li =0.36$ while the pitching amplitude and phase synchrony vary over the ranges of $7.5^{\circ} \leq \theta _0 \leq 30^{\circ}$ and $0 \leq \phi \leq 2\pi$, respectively (see varying $\theta_0$ in Table \ref{table:parameter}). Figure \ref{fig:varyamp}a presents the stable gap distances normalized by the nominal wake wavelength as a function of the phase difference. Unlike the stable gap distances with varying $Li$, the nominal wake wavelength can provide a good collapse with varying amplitude. In Figure \ref{fig:varyamp}b, the stable gap distances normalized by the actual wake wavelength are presented including the data for both varying Lighthill number and varying amplitude. It is found that the actual wake wavelength can provide an improved collapse of the data over the nominal wavelength especially when considering variations in both the pitching amplitude \textit{and} Lighthill number. Also, a linear relationship between the phase difference and the normalized stable gap distances, 
\begin{align}
\frac{G_{s}}{\lambda} = \frac{\phi}{2\pi} -0.29,
\end{align}
is observed, which is represented by the black dashed lines and determined by linear regression with the data. Figure \ref{fig:varyamp}c shows that $\lambda/\lambda_{0}$ maintains a nearly constant ratio with variations in amplitude, which is why there is a good collapse of the stable gap distances across a range of amplitudes regardless of whether $\lambda$ or $\lambda_0$ is used as a scaling factor.  Figure \ref{fig:varyamp}d presents the normalized collective efficiency as a function of the normalized stable gap distance.  The maximum efficiency enhancement is achieved for $G_s/\lambda<0.7$ for all pitching amplitudes just as in the varying Lighthill number data. As observed previously, peaks in efficiency enhancement occur at $\phi \approx \pi$ and troughs occur at $\phi \approx 0$, except for $\theta_{0} = 30^{\circ}$ where those trends are flipped. Benefits in propulsive efficiency of in-line schooling ($\eta_{c}^*>1$) are observed at nearly all stable gap distances over the range of $7.5^{\circ} \leq \theta_{0} \leq 22.5^{\circ}$ with $\theta_{0} = 15^{\circ}$ giving the global peak efficiency benefit when $G_s/\lambda<0.7$ and $\theta_{0} = 22.5^{\circ}$ giving the local peak efficiency benefits when $G_s/\lambda > 0.7$. In contrast, there is an efficiency penalty  at the pitching amplitude of $30^{\circ}$. 


\subsection{Scaling the wake wavelength}\label{sec:scaling}
\begin{figure}
    \centering
    \includegraphics[width=\textwidth]{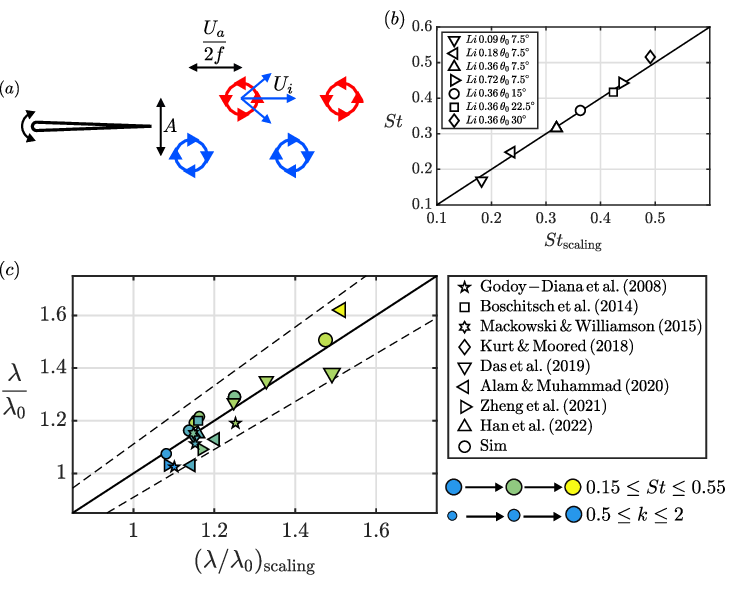}\vspace{0.in}
    \vspace{-0.1in}
    \caption{Proposed wavelength scaling law collapses data from two-dimensional experimental and numerical studies over a wide range of $St$ and $k$. $(a)$ Schematic of a purely pitching foil and its wake vortices. $(b)$ The Strouhal number from unconstrained pitching foil simulations compared with the scaling relation (\ref{eq15}). $(c)$ Wavelength data compared with the scaling relation (\ref{eq8}). Dashed lines represent $\pm 10$\% error. These data cover unconstrained simulations from the current study as well as simulations and experiments from previous constrained studies for $200 \leq Re < \infty$ \citep{Godoy-Diana2008, Boschitsch2014, Mackowski2015, Kurt2018, Das2019, Alam2020, Zheng2021, Han2022a}. The marker color from blue to yellow represents increasing $St$ while the marker size from small to large represents increasing $k$. Note that for the current unconstrained simulations first $St$ and $k$ are calculated with the scaling relations (\ref{eq15}) and (\ref{eq16}), which are then used in scaling relation (\ref{eq8}) to find $\lambda/\lambda_{0}$.}
   \label{fig:scaling}
\end{figure}

As shown in the previous sections, stable gap distances of in-line schooling scale with the wake wavelength, $\lambda$, of the isolated leader rather than the nominal wake wavelength, $\lambda_{0}$. Here, we present scaling laws to predict the wake wavelength from the Strouhal number and reduced frequency of the leader. In fact, the nominal wake wavelength is already a function of the reduced frequency,
\begin{gather}
\lambda_{0} = \frac{c}{k},\label{eq1}
\end{gather}
Therefore, we only need to provide a scaling law for the ratio of the wavelength over the nominal wavelength, which is equivalent to the speed of wake vortices relative to the nominal freestream speed,
\begin{gather}
\frac{\lambda}{\lambda_{0}} = \frac{U_{a}/f}{U/f} = \frac{U_{a}}{U},\label{eq2}
\end{gather}
where $U_{a}$ is the velocity of a wake vortex advecting downstream. This is calculated as
\begin{gather}
U_{a} = U + U_{i},\label{eq3}
\end{gather}
where $U_{i}$ is the induced streamwise velocity acting on a wake vortex from other wake vortices, which is approximated as the velocity induced by the two nearest vortices  (Figure \ref{fig:scaling}a). Therefore, the ratio of the wavelength over the nominal wavelength becomes
\begin{gather}
\frac{\lambda}{\lambda_{0}} = 1 + U^*_{i}, \label{eq4}
\end{gather}
where $U^*_{i} = U_i/U$. Based on the Biot-Savart law, the induced velocity is calculated as, 
\begin{gather}
U_{i} =  \frac{\Gamma A}{\pi r^2} \quad \text{with} \quad r = \sqrt{\frac{U^2_{a}}{4f^2} + A^2},\label{eq5}
\end{gather}
where $\Gamma$ is the strength of the wake vortices, and $r$ is the distance from the neighboring vortices to the vortex of interest. Solving equation (\ref{eq5}) gives (see appendix \ref{inducedscal} for details)
\begin{gather}
U^*_{i} = c_{1}\, k\,St. \label{eq7}
\end{gather}
where $c_1$ is a constant to be determined. We then have,
\begin{gather}
\frac{\lambda}{\lambda_{0}} = 1 + c_{1}\,k\, St. \label{eq8}
\end{gather}

The scaling law in equation (\ref{eq8}) can be applied to both constrained and unconstrained pitching foils. However, the Strouhal number and reduced frequency of unconstrained foils are not known \textit{a priori} since they depend on the mean swimming speed of the foils. We can provide an additional scaling law to predict these dimensionless variables as a function of the dimensionless input variables known \textit{a priori}, namely, the dimensionless amplitude and Lighthill number. To do this, we begin with a scaling law for the thrust of pitching foils developed in \cite{moored2018inviscid}, 
\begin{gather}
C_{T}^a = c_{2}\phi_{1} + c_{3}\phi_{2} + c_{4}\phi_{3}, \label{eq9}
\\ \phi_{1} = 1, \label{eq10}
\\ \phi_{2} = w(k) = \frac{3F}{2} - \frac{G}{2\pi k} + \frac{F}{\pi^2 k^2} - (F^2 + G^2)\left(\frac{1}{\pi^2k^2} + \frac{9}{4}\right), \label{eq11}
\\ \phi_{3} = A^*,
\end{gather}
where $C^a_{T} = \overline{T}/(\rho c s f^2 A^2)$ is the thrust normalized by the characteristic added mass force of a pitching foil,  $w(k)$ is the wake function, $F$ and $G$ are the real and imaginary components of Theodorsen's lift deficiency function, respectively \citep{Theodorsen1935, Garrick1936}, and $c_{2}$, $c_{3}$, and $c_{4}$ are constants to be determined. Under a high-frequency approximation, we have
\begin{gather}
k \rightarrow \infty, \quad \text{where} \; \; F = \frac{1}{2} \; \; \text{and} \;\; G=0. \label{eq12}
\end{gather}
Therefore, equation (\ref{eq9}) becomes,
\begin{gather}
C_{T}^{a} = c_{2} + c_{3}\left(\frac{3}{16} + \frac{1}{4\pi^2k^2}\right) + c_{4} A^*, \label{eq13}
\end{gather}
and since
\begin{gather}
St^2 = \frac{Li}{2C_{T}^a} \;\; \text{and} \;\; k=\frac{St}{A^*}\label{eq14},
\end{gather}
we have (see appendix \ref{standkscal} for details),
\begin{gather}
St = \sqrt{c_{2} Li + c_{3} {A^*}^2}\label{eq15}
\end{gather}
and 
\begin{gather}
k = \frac{1}{A^*}\sqrt{c_{2} Li + c_{3} {A^*}^2}.\label{eq16}
\end{gather}

Using the free-swimming data in Sections \ref{sec:Li} and \ref{sec:amp}, the coefficients in equations (\ref{eq15}) and (\ref{eq16}) are determined to be $c_{2} = 0.25$ and $c_3 = 0.15$, respectively by minimizing the squared residuals. Figure \ref{fig:scaling}b compares the data of the Strouhal number with the prediction of the scaling law. It is shown that the Strouhal number of unconstrained pitching foils can be well predicted by the scaling law. Here, the comparison of the reduced frequency with its prediction is not shown since it is simply the Strouhal number divided by the dimensionless amplitude, which produces a graph showing the same agreement.

Compiling the free-swimming data in the present study with data from previous numerical and experimental two-dimensional studies of constrained foils representing a wide Reynolds number range of $200 \leq Re <\infty$,
the coefficient in equation (\ref{eq8}) is then determined to be $c_{1} = 0.64$ by minimizing the squared residuals. Figure \ref{fig:scaling}c compares the ratio of the wavelength over the nominal wavelength with the scaling law prediction. The scaling law can predict the ratio of the wavelength over the nominal wavelength to within $\pm 10\%$ error. Note that this $10\%$ error also applies to the prediction of the wavelength since the wavelength is simply $\lambda/\lambda_{0}$ multiplied by $c/k$. Moreover, it is discovered that the wake wavelength can be up to $1.6$ times the nominal wake wavelength, indicating the inaccuracy in predicting the spacing of wake vortices with the nominal wake wavelength.

From equations (\ref{eq1}) and (\ref{eq8}), the scaling law of the wake wavelength can be explicitly written as
\begin{gather}
\lambda = c \left(\frac{1}{k} + c_{1}St\right). \label{eq17}
\end{gather}
This scaling law can be used for both constrained and unconstrained studies. For unconstrained studies, in which the Strouhal number and reduced frequency are not known \textit{a priori}, the scaling laws in equations (\ref{eq15}) and (\ref{eq16}) can be employed to predict the Strouhal number and reduced frequency.

\subsection{Enhancing efficiency in in-line stable formations by mismatched amplitudes}\label{sec:unmatched}

\begin{figure}
    \vspace{0.1in}    
    \centering
    \includegraphics[width=\textwidth]{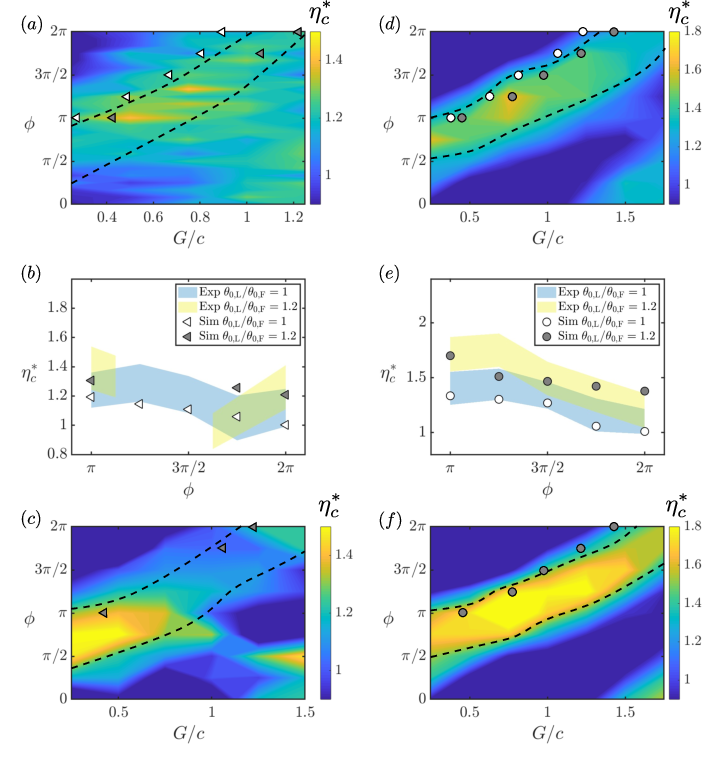}\vspace{0.0in}
    \vspace{-0.1in}
    \caption{Further efficiency enhancement  at in-line stable positions is achieved by increasing the leader's amplitude. The follower's amplitude is fixed at $7.5^{\circ}$, and the Lighthill number of the foils is fixed at $0.18$: $(a)$ The experimental efficiency contour plot that has the same kinematics as the free-swimming simulations with $\theta_{0,L}/\theta_{0,F}=1$.  White markers represent the stable positions of the free-swimming simulations with $\theta_{0,L}/\theta_{0,F}=1$ , and gray markers represent the stable positions of the free-swimming simulations with $\theta_{0,L}/\theta_{0,F}=1.2$. The high-efficiency zone is outlined by dashed lines. This zone is defined as the region where the efficiency is at or above $90\%$ of its maximum value for each phase.  $(b)$ $\eta_{c}^*$ at the stable positions as a function of $\phi$ from simulations (markers) and experiments (shaded areas). The upper and bottom limits of the shaded area represent the mean vaule plus and minus the standard deviation, respectively. $(c)$ The experimental efficiency contour plot that has the same kinematics as the free-swimming simulations with $\theta_{0,L}/\theta_{0,F}=1.2$.  The follower's amplitude is fixed at $15^{\circ}$, and  the Lighthill number of the foils is fixed at $0.36$: $(d)$ The experimental contour plot that has the same kinematics as the free-swimming simulations with $\theta_{0,F}/\theta_{0,L} =1$. $(e)$ $\eta_{c}^*$ at the stable positions as a function of $\phi$ from simulations and experiments. $(f)$ The experimental efficiency contour plot that has the same kinematics as the free-swimming simulations with $\theta_{0,F}/\theta_{0,L} =1.2$.}
   \label{fig:unmatched}
\end{figure}  

In the previous sections the physics of freely swimming leader-follower foil pairs were detailed when they used identical amplitudes of motion. In this section we turn our attention to probing the potential of \textit{mismatched} amplitudes, where the leader and follower use different amplitudes, to further improve the efficiency performance of the pairs while maintaining stable in-line formations. New unconstrained simulations (Table \ref{table:parameter}; varying $\theta_{0,L}/\theta_{0,F}$) and new constrained experiments (Table~\ref{table:parameterEXP}) are presented in this section. We are motivated to investigate mismatched amplitudes in response to the comparison of the free-swimming data from the current study with previous constrained measurements \cite[]{Kurt2018}.
\begin{table}
\centering
\begin{tabular}{ccccccccccccccc}
  & & $U_\infty$ & & & $f$ & & &     $\theta_{0,L}$& &$\theta_{0,F}$\\
Exp1 & & $0.071$ m/s & & &   $0.75$ & & &    $7.5^\circ$ & &$7.5^\circ$\\
Exp2 & & $0.092$ m/s & & &   $0.75$  & & & $9^\circ$ & & $7.5^\circ$\\
Exp3 & & $0.103$ m/s & & &   $0.75$  & & & $15^{\circ}$& & $15^{\circ}$\\
Exp4 & & $0.117$ m/s  & & &  $0.75$  & & & $18^{\circ}$& & $15^{\circ}$\\
\end{tabular}
\caption{\label{table:parameterEXP} Experimental variables used in the study. Experiments $1$--$4$ correspond to Figure \ref{fig:unmatched}a, c, d, and f, respectively.}
\end{table}

To illustrate this motivation, Figure \ref{fig:unmatched}a presents previous constrained experimental data \cite[]{Kurt2018} of the normalized efficiency as a function of $G/c$ and $\phi$ when $U=0.071 \, \text{m}/\text{s}$, $f = 0.75\, \text{Hz}$, and $\theta_{0,L} = \theta_{0,F} = 7.5^{\circ}$. A high-efficiency zone (outlined by dashed lines) where efficiency schooling benefits are maximized is observed. This zone is defined as the region where the efficiency is at or above $90\%$ of its maximum value for each phase. The foil kinematics/geometry used to produce this contour are precisely the same as those used in the free-swimming simulations with $Li = 0.18$ and $\theta_{0}=7.5^{\circ}$ presented in Section \ref{sec:Li}, so we directly compare the data by plotting the stable gap distances found in the free-swimming simulations (white markers) on top of the experimental contour in Figure \ref{fig:unmatched}a. It is discovered that the stable positions of freely swimming foils, while receiving some efficiency benefit, do not lie in the high-efficiency zone, but rather near its edge. In this way, the constrained experiments act as a map to show that there are potential additional efficiency benefits that could be unlocked if the foils' stable gap distance were larger at a given phase difference. To achieve this, the key idea to understand is that the high efficiency benefits are driven by high follower thrust benefits \cite[]{Kurt2018}, and if the follower is at a gap distance in the high efficiency zone then it receives too much additional thrust such that it will swim faster than the leader and reduce the gap distance until their thrusts are equilibrated. We hypothesize that by increasing the leader's amplitude relative to the follower (or reducing the follower's amplitude relative to the leader) the high thrust of the follower in the high efficiency zone can be matched by the leader thereby extending the stable gap distance to reside in that zone. Then the foil pair may be able to maximize its efficiency while maintaining a stable formation.

To test out this hypothesis, new free-swimming simulations were conducted where the leader-to-follower amplitude ratio is increased to $\theta_{0,L}/\theta_{0,F} = 1.2$. It's revealed that the stable gap distances from these simulations (gray markers in Figure \ref{fig:unmatched}a) are indeed extended at the same phase difference, and they \textit{seem} to reside in the high-efficiency zone. Note that at $\phi = 5\pi/4$ and $3\pi/2$ the stable formations are lost. Figure \ref{fig:unmatched}b compares the efficiency benefit calculated from the free-swimming simulations at the stable gap distances of matched and mismatched amplitudes and it is also observed that the efficiency \textit{is} increased for the mismatched amplitude case, as hypothesized. In addition, the increased efficiency benefit is observed in experimental data. Despite the extended stable gap distances and the increased efficiency, it is not yet clear that the foil pair resides in the high-efficiency zone, since the contour plot in Figure \ref{fig:unmatched}a represents the data from experiments on \textit{matched} amplitudes of motion. When there are mismatched amplitudes the high-efficiency zone itself may be altered. Therefore, Figure \ref{fig:unmatched}c presents data from \textit{new experiments} of foil pairs with mismatched amplitudes ($\theta_{0,L}/\theta_{0,F} = 1.2$) with the stable gap distances of the free-swimming simulations plotted on top. Now, it is confirmed that the stable positions are indeed pushed into the high-efficiency zone when the amplitude ratio is increased to $\theta_{0,L}/\theta_{0,F} = 1.2$. Surprisingly, the efficiency benefits in the zone and its neighborhood are also increased above the case of matched amplitudes of motion.

Now, the matched amplitude case presented in Figure \ref{fig:unmatched}a--\ref{fig:unmatched}c ($\theta_0 = 7.5^{\circ}$) is not the baseline case that maximizes the efficiency, so does this approach continue to increase the efficiency benefits even for the maximum efficiency baseline case, that is, $\theta_{0} =15^{\circ}$ and $Li=0.36$? To address this question, Figure \ref{fig:unmatched}d and \ref{fig:unmatched}f present new constrained experimental data of the normalized efficiency for $\theta_{0,L}/\theta_{0,F} = 1$ and $\theta_{0,L}/\theta_{0,F} = 1.2$, respectively, with $\theta_{0,F} =15^{\circ}$ and $Li=0.36$. Figure \ref{fig:unmatched}e presents the efficiency enhancement calculated from the simulations of the matched and mismatched amplitudes when the foils are freely swimming at their stable gap distances. The stable positions from the free-swimming simulations with matched and mismatched amplitudes are represented by the white and gray markers, respectively. It is discovered that the stable positions for matched amplitudes are still near the edge of the high-efficiency zone (Figure \ref{fig:unmatched}d) and that by increasing the leader's amplitude from $\theta_0 = 15^{\circ}$ to $18^{\circ}$ it does increase the foils' stable gap distance, and their efficiency enhancement increases from a peak baseline value of a $33\%$ improvement over isolated foils to more than doubling that to a $70\%$ improvement over isolated foils (Figure \ref{fig:unmatched}e). Furthermore, this increased efficiency benefit revealed by the simulations agrees well with the experimental data in Figure \ref{fig:unmatched}e.  Despite the substantial improvement in efficiency and the increase in the foils' stable gap distance, the foils still do not reside in the center of the high efficiency zone (Figure \ref{fig:unmatched}f), but rather stay near its edge.  It is observed that the zone shifts to larger gap distances at the same phase difference and the substantial efficiency enhancement over the matched amplitude case is driven by a rise in the collective efficiency levels for these mismatched amplitudes of motion. 

The constrained-foil measurements presented in Figure \ref{fig:unmatched}, show that the general feature of high efficiency bands observed in the matched amplitude maps are reproduced in the mismatched amplitude maps regardless of changes in the efficiency levels. This is not surprising since the bands are indicative of a vortex-body impingement mechanism \cite[]{Boschitsch2014}, which should not fundamentally change with mismatched amplitudes. In this way, constrained measurements of matched amplitudes can provide a map that can guide the choice of kinematics of free-swimming hydrofoils towards high efficiency. Instead of mismatched amplitudes, other strategies for tailoring stable formations could be used and investigated such as mismatched frequencies, non-sinusoidal motions, and intermittent swimming. The findings in this study can aid in the design of high-efficiency bio-robots that school in formation. Future work should examine data of fish schools to determine whether mismatched amplitudes is a strategy used in biology to maximize efficiency.

\section{Conclusions}
In this study, we simulate two pitching hydrofoils freely swimming in an in-line arrangement. It is found that self-organizing in-line stable formations with efficiency benefits exist under a wide range of flow conditions defined by the phase synchrony, pitching amplitude, and Lighthill number. When leading-edge flow separation is suppressed, the follower always collides into the leader, and in-line stable formations cannot be formed regardless of the kinematics. Additionally, the characteristic vortex pairs above and below the follower that form during in-line schooling disappear when there is no leading-edge flow separation. Comparing constrained simulations with and without leading-edge flow separation reveals that separation acts a dynamic drag source on the follower that varies with the synchrony and gap spacing. In fact, increasing the drag of the follower by giving it a larger Lighthill number can be enough to form in-line stable formations, however, this approach only qualitatively agrees, not quantitatively to simulations and experiments where leading-edge separation on the follower is present. 

From free-swimming simulations over a wide range of flow conditions typical of biology it is discovered that the nominal wake wavelength widely used in previous work, does not scale the in-line stable positions, particularly for variations in the Lighthill number. The nominal wavelength can achieve a good collapse for variations in the amplitude, since the actual-to-nominal wavelength ratio is nearly constant across pitching amplitudes. However, this ratio exhibits a large variation across different Lighthill numbers, where the wake wavelength is discovered to be up to $1.6$ times the nominal wake wavelength, highlighting the inaccuracy in using the nominal wavelength to predict the spacing of wake vortices. When using the actual wake wavelength a good collapse occurs for variations in phase, amplitude, and Lighthill number, and reveal a linear relationship between the phase difference and the normalized gap distance. By considering the velocity induced on a wake vortex by its neighbours, a scaling law of the wake wavelength is developed. This scaling law is applied to both unconstrained and constrained data from the current study and previous studies, showing good agreement to within $\pm 10\%$ error across a wide Reynolds number range of $200 \leq Re <\infty$. 

Considering the efficiency performance of schooling in-line foils at stable gap distances, the maximum efficiency enhancement is achieved for compact arrangements where $\lambda<0.7$ and with the phase difference of $\phi = \pi$ for all amplitudes except for $\theta_{0}=30^{\circ}$ where there is, in fact, an efficiency penalty compared to an isolated foil. The maximum efficiency benefit also occurs at the lowest Lighthill numbers examined of $Li = 0.09$ and at the optimal amplitude of $\theta_{0} = 15^{\circ}$.

By using measurements of matched amplitude constrained foils as a guide, it was hypothesized that by increasing the amplitude of the leader relative to the follower -- using mismatched amplitudes -- the pair of foils could self-organize with a gap spacing in a zone of high efficiency. New constrained experiments and free-swimming simulations confirmed this hypothesis and showed that by using matched amplitudes the peak efficiency of the foil pair could increase by 33\% over isolated foils while using mismatched amplitudes could more than double that to a 70\% increase over isolated foils. This occurs for mismatched amplitudes by (1) increasing the stable gap distance between foils such that they are pushed into a high-efficiency zone, and (2) by raising the level of efficiency in these zones. This study not only presents a way to tailor in-line stable formations towards high efficiency via mismatched amplitudes, but also bridges the gap between constrained and unconstrained studies. It's shown that constrained foil measurements can be used as a map of the potential benefits of schooling, which can guide the choice of kinematics of free-swimming hydrofoils towards high efficiency.

\section*{Acknowledgments}
This work was supported by the Office of Naval Research under Program Director Dr. Robert Brizzolara on MURI grant number N00014-22-1-2616.

\section*{Declaration of interests}
The authors report no conflict of interest.

\appendix

\section{Derivation of scaling laws}\label{app}

\subsection{Scaling law for the induced velocity} \label{inducedscal}
From equations (\ref{eq3}) and (\ref{eq5}), we have
\begin{gather}
U_{i} = \frac{\Gamma A}{\pi\left(\frac{(U+U_{i})^2}{4f^2} + A^2\right)}. \label{eqa1}
\end{gather}
After non-dimensionalization, equation (\ref{eqa1}) becomes
\begin{gather}
U^*_{i} = \frac{\Gamma^*}{\pi A^{*} U\left(\frac{(1+U^*_{i})^2}{4St^2} + 1\right)}, \label{eqa2}
\end{gather}
where $\Gamma^*$ is the non-dimensional wake circulation, $\Gamma^* = \frac{\Gamma}{Uc}$.
After organization equation (\ref{eqa2}) becomes 
\begin{gather}
{U^*_{i}}^3 + 2{U^*_{i}}^2 + U^*_{i}(1+4St^2) = \frac{4St^2\Gamma^*}{\pi A^*}. \label{eqa3}
\end{gather}
Since $U^*_{i}$ and $St$ are less than $1$, we neglect the terms of ${U^*_{i}}^3$ and $U^*_{i}St^2$ in equation (\ref{eqa3}) for simplification, which gives 
\begin{gather}
2{U^*_{i}}^2 + U^*_{i} = \frac{4St^2\Gamma^*}{\pi A^*}. \label{eqa4}
\end{gather}
Solving equation (\ref{eqa4}) gives
\begin{gather}
U^*_{i} = -\frac{1}{4} + \frac{1}{4}\sqrt{1+\frac{32St^2\Gamma^*}{\pi A^*}}. \label{eqa5}
\end{gather}
Following \cite{Han2024} and \cite{moored2018inviscid}, we have 
\begin{gather}
\Gamma^* \propto \Gamma^*_{0}\frac{k^*}{1+k^*}, \label{eqa6}
\end{gather}
with
\begin{gather}
k^* = \frac{k}{1+4St^2} \;\; \text{and} \;\;\Gamma^*_{0} \propto \frac{3\pi^2 St}{4}, \label{eqa7}
\end{gather}
where $\Gamma^*_{0}$ is the non-dimensional quasi-steady circulation of a foil pitching about its leading edge. Then we have
\begin{gather}
\frac{St^2\Gamma^*}{\pi A^*} \propto \frac{k^2St^2}{1+4St^2+k}. \label{eqa8}
\end{gather}
So equation (\ref{eqa5}) becomes
\begin{gather}
U^*_{i} = -\frac{1}{4}  + \frac{1}{4}\sqrt{1+c_1 \left(\frac{ k^2St^2}{ 1+4St^2+k}\right)}, \label{eqa9}
\end{gather}
where $c_{1}$ needs to be determined. Since the last term in equation (\ref{eqa9}) is much larger than the other terms, equation (\ref{eqa9}) can be approximated as
\begin{gather}
U^*_{i} \approx c_{1}\frac{ k\, St}{ \sqrt{1+4St^2+k}}. \label{eqa10}
\end{gather}
Furthermore, since the variation of the denominator is much less than the numerator in equation (\ref{eqa10}), we approximate the denominator as a constant and simplify the equation, which gives
\begin{gather}
U^*_{i} = c_{1}\, k\, St.\label{eqa11}
\end{gather}

\subsection{Scaling law for the Strouhal number and reduced frequency of unconstrained foils} \label{standkscal}

From equation (\ref{eq13}) and (\ref{eq14}), we have
\begin{gather}
St^2  = \frac{Li}{2\left[c_{2} + c_{3}\left( \frac{3}{16} + \frac{{A^*}^2}{4\pi^2 St^2} \right) + c_{4} A^*\right]} \label{eqa12}.
\end{gather}
After organization we have
\begin{gather}
St^2 \left(c_{2} + \frac{3}{16} c_{3} + c_{4} A^* \right) = \frac{Li}{2} - \frac{c_{3} {A^*}^2}{4\pi^2}\label{eqa13}.
\end{gather}
Therefore, the Strouhal number is calculated as
\begin{gather}
St = \sqrt{ \frac{\frac{Li}{2} - \frac{c_3 {A^*}^2}{4 \pi^2}}{c_{2} + \frac{3}{16}c_{3} + c_{4} A^*}    }.\label{eqa14}
\end{gather}
Because the variation of the denominator is much less than the numerator in equation (\ref{eqa14}), we approximate the denominator as a constant and simplify equation (\ref{eqa14}) to 
\begin{gather}
St = \sqrt{c_{2} Li + c_{3} {A^*}^2}.\label{eqa15}
\end{gather}
In addition, the reduced frequency is calculated as
\begin{gather}
k = \frac{St}{A^*} = \frac{1}{A^*}\sqrt{c_{2} Li + c_{3} {A^*}^2}.\label{eqa16}
\end{gather}

\FloatBarrier
\bibliography{MAIN.bib}
\bibliographystyle{jfm}

\end{document}